\documentclass{svproc}
\usepackage{url}
\usepackage{graphicx}
\usepackage{booktabs}
\usepackage{subfig}
\usepackage{wrapfig}
\usepackage{amsmath}
\usepackage[para]{footmisc}

\begin{document}
\mainmatter              % start of a contribution
\title{A Generic Scalable Method for Scheduling Distributed Energy Resources using Parallelized Population-based Metaheuristics}
\titlerunning{A Scalable Method for Scheduling Distributed Energy Resources using EAs}  % abbreviated title (for running head)

	\author{Hatem Khalloof\and
	Wilfried Jakob\and
	Shadi Shahoud \and
	Clemens Duepmeier \and
	Veit Hagenmeyer}
\authorrunning{H. Khalloof et al.}
%\\ Hermann-von-Helmholtz-Platz 1, \\ 76344 Eggenstein-Leopoldshafen,
\institute{Institute of Automation and Applied Informatics (IAI) \\ Karlsruhe Institute of Technology (KIT), Karlsruhe, Germany \\
	%\url{www.iai.kit.edu}
	\email {\{hatem.khalloof,wilfried.jakob,shadi.shahoud,clemens.duepmeier,
		veit.hagenmeyer\}@kit.edu}
}

\maketitle              % typeset the title of the contribution

\begin{abstract}
Recent years have seen an increasing integration of distributed renewable energy resources into existing electric power grids. Due to the uncertain nature of renewable energy resources, network operators are faced with new challenges in balancing load and generation. In order to meet the new requirements, intelligent distributed energy resource plants can be used which provide as virtual power plants e.g. demand side management or flexible generation. However, the calculation of an adequate schedule for the unit commitment of such distributed energy resources is a complex optimization problem which is typically too complex for standard optimization algorithms if large numbers of distributed energy resources are considered. For solving such complex optimization tasks, population-based metaheuristics --as e.g. evolutionary algorithms-- represent powerful alternatives. Admittedly, evolutionary algorithms do require lots of computational power for solving such problems in a timely manner. One promising solution for this performance problem is the parallelization of the usually time-consuming evaluation of alternative solutions.
In the present paper, a new generic and highly scalable parallel method for unit commitment of distributed energy resources using metaheuristic algorithms is presented. It is based on microservices, container virtualization and the publish/subscribe messaging paradigm for scheduling distributed energy resources. Scalability and applicability of the proposed solution are evaluated by performing parallelized optimizations in a big data environment for three distinct distributed energy resource scheduling scenarios. 
Thereby, unlike all other optimization methods in the literature --to the best knowledge of the authors--, the new method provides cluster or cloud parallelizability and is able to deal with a comparably large number of distributed energy resources.  The application of the new proposed method results in very good performance for scaling up optimization speed.

\end{abstract}
\section{Introduction}
Renewable Energy Resources (RERs) are recently widely integrated into the grid paving the road for more clean and environment-friendly energy. To facilitate the adoption and management of such RERs, the transition from a traditional centralized grid (macrogrid) to  more decentralized grids (microgrids) is required \cite{ref1Padiyar,ref2IEEE}.
Microgrids encompass respectively a localized group of Distributed Energy Resources (DERs) where each DER represents a small or larger scale and self-autonomous sub-system connected to an electricity network. DERs provide renewable energy generation and/or improve the overall power system reliability by balancing the energy supply and demand in a specific part of a power network by providing flexible load options or storage. Typically, a DER encompasses a group of small generation units such as PVs, wind turbines and diesel generators, electrical loads (demand-response) e.g. electric vehicles or flexible heating systems, and maybe storage. DERs interconnect bidirectionally to the grid through one or more Point(s) of Common Coupling (PCC) \cite{ref1DER}. By the time, the usage of DERs in smart grids will dramatically increase providing more clean energy generated from RERs and additionally also maintaining and increasing power quality and system reliability. The flexibility of microgrids provides a significant potential to promote and integrate more DERs for featuring their beneficial traits.
%Microgrids encompass respectively a localized group of Distributed Energy Resources (DERs) as e.g. batteries, fuel generators, solar, wind and gas turbines and distributed electrical loads with demand-response. The flexibility of microgrids provides a significant potential to promote and integrate more DERs to reduce greenhouse gases and meet the increasing energy demand.
Despite being highly effective, microgrids have some limitations such as lack of system protection and customer privacy. Moreover, by increasing the number of DERs in the grid and due to the uncertainties of RERs and load, the efficient control and optimal usage of DERs by finding the proper schedule for using them represents a big challenge \cite{ref4microgrid}. 

In general, scheduling problems e.g. scheduling DERs is an NP-hard optimization problem and therefore is typically too complex to be solved by exact optimization methods, especially if large size optimization problems are considered \cite{NpHardDER,NpHard}. Metaheuristics such as Evolutionary Algorithms (EAs) became one of the most robust methods to solve such complex problems by finding good local optima or even the global one.
%Hence,  developing new concepts, optimization approaches and software services based on modern technologies and artificial intelligence optimization techniques e.g.  EAs are promising approaches for solving such complex optimization problems  as scheduling of DERs.  %to  calculate the provision of flexible DERs to energy grids
The central concept of EAs is a population consisting of individuals representing tentative solutions. The individuals encode possible solutions and they are used to explore many areas of the solution space in parallel. Each individual is evaluated by a fitness function to identify its suitability as a solution for the problem. Genetic operators, namely recombination and mutation, are iteratively applied to individuals to generate a new offspring for each generation until a termination criterion has been reached \cite{ref5EA,ref6EA}. This approach of having a population of solutions and evaluating them over and over again takes a lot of computational resources for large problem sizes. Therefore, applying EAs for large scale optimization and NP-hard problems such as the problem of scheduling a large number of DERs can be time-consuming and computationally expensive. %Moreover, DERs can contain several heterogeneous energy resources resulting in a more complex and  heterogeneous search space.  
To speed up EAs, three different parallelization models, namely the Global Model (Master-Slave Model), the  Fine-Grained Model and the Coarse-Grained Model have been introduced and investigated in \cite{ref7EA}. In the Global Model, the evaluation step is parallelized over several computing units  (called slaves). In the  Fine and Coarse-Grained Models, the population is structured to apply the genetic operators in parallel.

Over the last decades, various approaches and frameworks e.g.
\cite{ref55PEA,ref5PEA,ref56PEA,ref69PEA,ref57PEA,ref82PEA,ref80PEA,ref65PEA,ref62PEA,ref2PEA,ref85PEA,ref83PEA,ref84PEA,ref78PEA,ref7PEA,ref71PEA,ref6PEA}
have been introduced to enable the parallel processing of EAs following the above three parallelization models. For most of these frameworks e.g.
\cite{ref55PEA,ref5PEA,ref56PEA,ref69PEA,ref57PEA,ref65PEA,ref62PEA,ref85PEA,ref7PEA,ref71PEA,ref6PEA}, a monolithic software architecture was the classical approach for the implementation which decreases the modularity, usability and maintainability of  the application. Recently, Big Data technologies such as Hadoop  and Spark  have been applied to speed up EAs e.g. \cite{jmetal,ref69PEA,ref65PEA,ref62PEA,ref78PEA,ref71PEA,ref6PEA}. However, most of these approaches also have a monolithic architecture which lacks hard boundaries and tends to become, with added functionality, complex and tightly coupled. This, in fact, limits the ability to provide simple and practical methods to plug in problem-specific functionality  e.g. simulators and even to integrate  existing EAs. By emerging modern software technologies, namely microservices, container virtualization and the publish/subscribe messaging paradigm, the parallelization of EAs in cluster and cloud environments to speed up EAs has become even more relevant, see e.g. \cite{ref82PEA,ref80PEA,ref78PEA,ref6PEA}.  Unlike monolithic applications, a microservices-based application contains several  small, autonomous, highly cohesive and decoupled services that work together to perform a specific task. Since all services are independent from each other, each microservice is able to utilize its own technology stack allowing great flexibility. The independence of the services allows each service to scale on demand. Microservice applications comprise two main features, namely modularity and technology heterogeneity which allow the microservices to be developed by different teams based on different technologies. These advantages combined with container runtime automation unlock the full potential of a parallelized EA by executing it on large scale computing clusters \cite{ref2PEA}.

In the present work, a new highly scalable, generic and distributed approach to schedule DERs is introduced. The microservice and container virtualization-based framework presented in \cite{ref2PEA} is adapted to carry out the required tasks. As the simulation based evaluation is by far the most time consuming part, the proposed framework distributes EAs according to the Global Model (Master-Slave model) \cite{ref7EA} where the evaluation is distributed over several computing units. On-demand deployment of services on a high performance distributed computing infrastructure, namely  a computing cluster, is supported. 
To validate the functionality of the proposed parallel approach, the EA GLEAM (General Learning Evolutionary Algorithm and Method) \cite{ref1GLEAM} is integrated into the framework. As a test task, the creation of an hourly day-ahead schedule plan for a simulated microgrid is chosen. In this microgrid, three use case scenarios are defined. In the first and second scenario, 50 DERs are considered to cover the required power for a simple load profile. 
In the third scenario 100 DERs are utilized to supply the requested  power for a more complex load profile. 
For evaluation of the scalability and the performance of the new solution, the framework is deployed on a cluster with four nodes, each one has 32 Intel cores (2,4 GHz).

The rest of the present paper is structured as follows. The next section reviews some related work for scheduling DERs based on EAs. Section \ref{sec:arch} introduces the extended architecture of the proposed approach. Section \ref{sec:gleam} starts with a short introduction of the EA GLEAM serving as metaheuristic. Section \ref{sec:eva} introduces a description of the defined use case scenarios, the deployment on a cluster and the obtained results. Section \ref{sec:cons} concludes with a summary and planned future work.
\section{Related Work}
\label{sec:relatedwork}
EAs have attracted the attention of researchers to solve several optimization problems in energy systems, namely expansion planning, e.g. \cite{ref2RW}, maintenance scheduling, e.g. \cite{ref4RW}, scheduling energy resources (unit commitment) and economic dispatch \cite{ref9RW,ref6RW,ref10RW,ref11RW,ref12RW,ref13RW,ref8RW,ref14RW,ref5RW,ref7RW}, to name a few. 
In recent extensive overviews, Zia et al. \cite{ref15RW} and Alvarado-Barrios et al. \cite{ref1EVA} presented  comprehensive studies about different methods and techniques used in Energy Management Systems (EMS) to optimize and schedule the operations. In the following sections, we summarize some of these works studied in \cite{ref1EVA} and \cite{ref15RW} focusing on using EAs (especially Genetic Algorithms GAs) for scheduling DERs.

For the problem of scheduling DERs, the authors of \cite{ref9RW,ref10RW,ref11RW,ref12RW,ref13RW,ref14RW} implemented GAs to schedule the power generation in microgrids. Several microgrids with  sizes ranging  from six to 12 DERs and a wide variety of generators e.g. PVs, wind turbines, microturbines and diesel engines and energy storage systems (batteries) are considered in these studies. While in  \cite{ref10RW,ref12RW,ref13RW,ref14RW} standard GA implementations were used, in \cite{ref9RW} a memory-based GA algorithm  and in \cite{ref11RW} an improved GA combined with simulated annealing technique were utilized to accelerate GAs for finding the optimal schedule. Minimizing the operation cost was the objective function for all these works. However, in \cite{ref12RW} the eco-pollutant treatment costs were additionally considered as objective function. Quan et al. in \cite{ref14RW} defined five  deterministic and four stochastic case studies solved by GA. They concluded that GA can introduce robust solutions for stochastic optimization problems. 

All the previous works focused on developing a respective new optimization algorithm using non-distributed EAs for achieving better solution quality. They tested their proposed solutions with microgrids consisting of small numbers of energy resources and deployed them using a monolithic software architecture. This limits the scalability and modularity of the proposed system which in turn restricts the possibility to handle scalable number of DERs. Despite the principally satisfactory performance of using EAs for scheduling DERs, there is no generic, parallel and scalable solution to facilitate the usage of the EAs for scheduling a scalable energy system on a scalable runtime environment such as a cluster and to work efficiently with other components e.g. forecasting frameworks and simulators.

Therefore, the present work introduces a highly parallel and scalable approach using a proven and established software environment based on microservices and container virtualization with full runtime automation on big computing clusters and an easy-to-use web-based management for scheduling  DERs based on distributed EAs. It provides a highly flexible environment for solving the problem of scheduling DERs for external applications e.g. EMS, and allows easy communication with other needed tools such as forecasting tools and external simulators.
\section{Microservice and Container Virtualization Approach for Scheduling DERs using Parallelized EAs}
\label{sec:arch}
In the following, the conceptual architecture of the proposed generic distributed approach for scheduling DERs based on EAs are detailed. The last subsection introduces GLEAM which is used as concrete EA for evaluating the approach. 
%%%%%%%%%%%%%%%%%%%%%%%%%%%%%%%%%%%%%%%%%%%%%%%%%%%%%%%%%%%%%%%%%%%%%%%%%%%%%%%%%%%%%%%%%%%%%%%%%%%%%%%%%%%%%%%
%\subsection{Distributed Energy Resources (DERs)}
%\label{sec:HDERS}
%A DER represents a small or larger scale and self-autonomous sub-system connected to an electricity network for providing renewable energy generation and/or improving the overall power system reliability by balancing the energy supply and demand in a specific part of a power network by providing flexible load options or storage. Typically, a DER encompasses a group of small generation units such as PVs, wind turbines and diesel generators, electrical loads (demand-response) e.g. electric vehicles or flexible heating systems, and maybe storage. 
%%A DER may also contain uncontrollable loads which complicate the DER coordination and management.
%DERs interconnect bidirectionally to the grid through one or more Point(s) of Common Coupling (PCC) \cite{ref1DER}.
%By the time, the usage of DERs in smart grids will dramatically increase providing more clean energy generated from RERs and additionally also maintaining and increasing power quality and system reliability. 
%%%%%%%%%%%%%%%%%%%%%%%%%%%%%%%%%%%%%%%%%%%%%%%%%%%%%%%%%%%%%%%%%%%%%%%%%%%%%%%%%%%%%%%%%%%%%%%%%%%%%%%%%%%%%%%%
\subsection{Microservice and Container Virtualization Approach}
\begin{figure}[!h]
	\centering
	\includegraphics[width=0.8\textwidth]{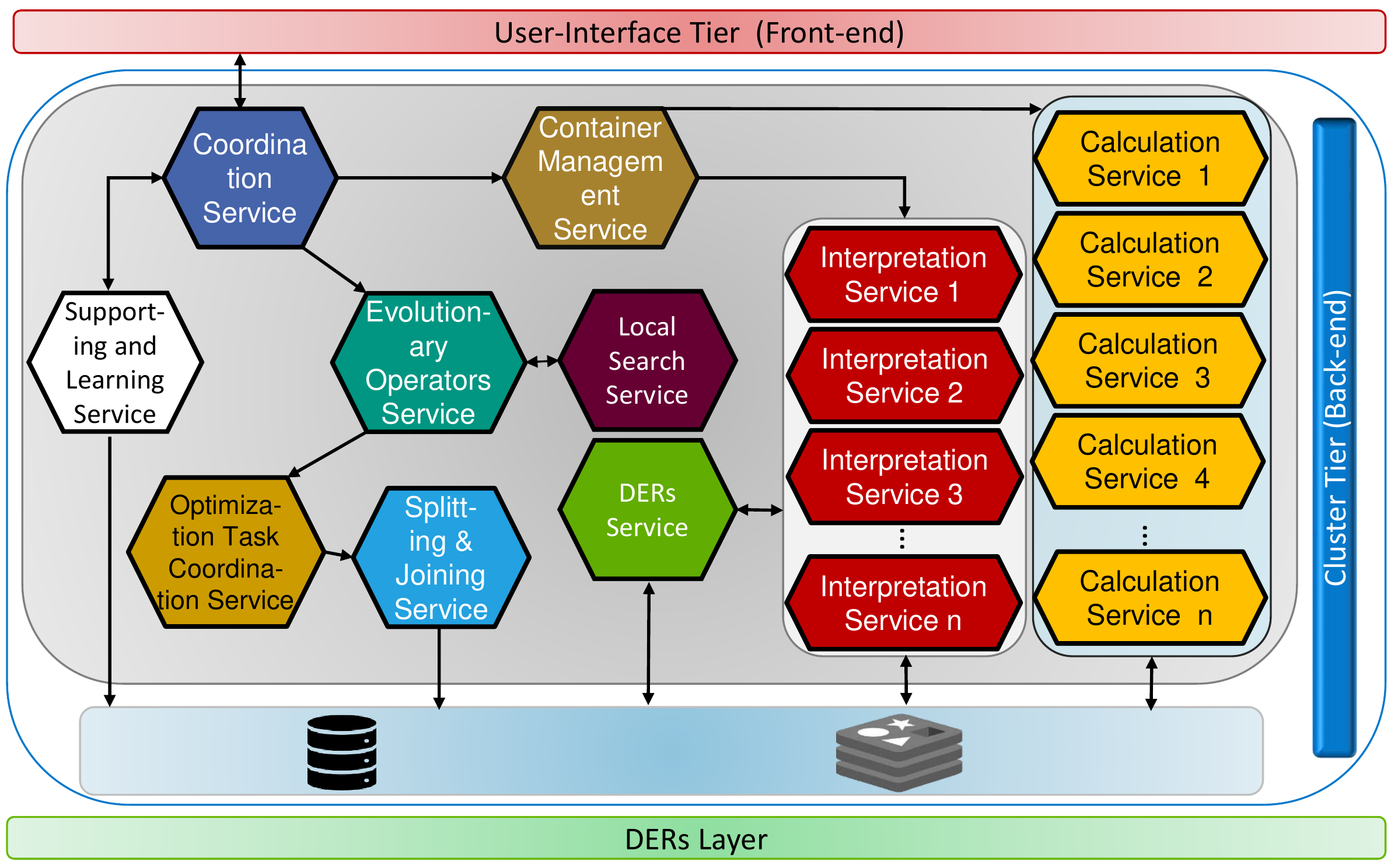}
	\caption{The conceptual architecture of the proposed architecture with detailed container layer}
	\label{fig:containerlayer}
\end{figure}

The conceptual architecture of the proposed highly scalable metaheuristic optimization solution is derived from \cite{ref2PEA}. As shown in Figure  \ref{fig:containerlayer}, the architecture has three main tiers, namely the User Interface (UI) Tier, the Cluster Tier and the Distributed Energy Resources (DERs) Tier. On the front-end, the UI Tier is dedicated to user interaction, e.g. input for defining optimization tasks, uploading optimization models, starting and stopping optimization tasks and presenting the obtained results. The UI Tier introduces a simple web-based UI to manage the interaction with the back-end tier.
\vspace{-5mm}
On the back-end, the Cluster Tier contains two sub-layers, namely the Container Layer and the Data Layer. The Container Layer contains all the services necessary to execute a parallel EA in the framework. This includes not only  the services that actually execute the EA, but also services for coordinating the execution and distributing the data. Each service is realized as a microservice running in a containerized environment. The Data Layer stores all data and acts as an intermediate for message exchange. To reflect the different properties of the data, the Data Layer is subdivided into a Persistent  and Temporary Storage (in-memory database). On the one hand, the Persistent Storage is responsible for storing the data needed for each optimization  task such as forecasting data and the final results for further usage. On the other hand, the Temporary Storage stores the intermediate data that are exchanged between services when performing an optimization job. The Temporary Storage also realizes a publish/subscribe message exchange pattern to improve the decoupling among the services. The DERs Tier contains abstractions of the DERs that have to be scheduled. Each DER abstraction provides required data, e.g. the predicted generation, consumption and market price for the period considered, and the necessary technical properties about the nonrenewable sources e.g. diesel generators. This data is needed by the services within the Container Layer for creating an optimized scheduling plan.
In the following sections, the container layer with the implemented services will be described in greater detail.

\subsubsection{Container Layer}
For finding the optimal scheduling plan for a group of DERs using parallel EAs, the software solution needs to perform several tasks, namely coordination of the execution of tasks, i.e. managing the containers, starting and managing external simulators and executing the parallel EAs for generating, splitting, distributing, and evaluating the chromosome lists, and collecting and joining the subresults to form the final results and applying the genetic operators. These tasks are performed by ten decoupled and cohesive microservices as shown in Figure \ref{fig:containerlayer}. The presented microservices are adapted from \cite{ref2PEA} for scheduling DERs using parallel EAs based on the global parallelization model.  Three new services, namely Supporting and Learning Service, DERs Service and Interpretation Service are added. Some of the existing microservices are modified and renamed to reflect their extended functionalities and new tasks. The Distribution \& Synchronization service is split into two services depending on its functionalities, namely the Optimization Task Coordination Service and the Splitting \& Joining Service. The framework is designed with this hierarchical structure for facilitating manageability and allowing extensibility. Adapted and newly added microservices are described in detail below.

{\textbf{Coordination Service}}
The Coordination Service  (formerly named Optimization Job Management Service \cite{ref2PEA}) is one of the core parts of the framework. It acts not only  as a coordinator for multiple jobs, but also for the whole framework.
After receiving the configuration, the Coordination Service asks the Container Management Service to start the required number of instances of the Interpretation Service and the Calculation Service. As soon as the required services are booted up, the Coordination Service calls the Evolutionary Operators Service to create the requested number of  chromosomes of the initial population. At the end of an optimization job, the Coordination Service receives the aggregated result and sends it to the visualization component to be visualized. The Coordination Service does not act as a master in the global model, rather it coordinates the services by initialization and termination. 
 
{\textbf{Evolutionary Operators Service}}
This service performs the task  of the master in the global model. At first, it generates the initial population when called by the Coordination Service. Then, it calculates the fitness function  to identify the individuals surviving for the next generation. Furthermore, it applies the genetic operators, namely crossover and mutation as well as the selection operation to generate the offspring. 

{\textbf{Optimization Task Coordination Service}}
The Optimization Task Coordination Service  (formerly named as Distribution \& Synchronization Service \cite{ref2PEA}) coordinates  one optimization task by e.g. assigning a Task ID, selecting one of the available simulation models that is available and starting and stopping an optimizing task. Indeed, it acts as a coordinator between the Evolutionary Operators Service  and other services.  

{\textbf{Splitting \& Joining Service}}
The Splitting \& Joining Service  (formerly named as Distribution \& Synchronization Service \cite{ref2PEA}) receives the offspring, i.e. the chromosome list from the Evolutionary Operators Service. Afterwards, it evenly splits and distributes the population to the Interpretation Service instances.  By finishing the distribution of the subpopulations successfully, the Interpretation Services start the interpretation processes by receiving a start signal from the Splitting \& Joining Service. As soon as the optimization task is finished, the Splitting \& Joining Service creates the overall result list matching the original list format supported by Evolutionary Operators Service by joining the partial results. Finally, the overall result is sent back to the Optimization Task Coordination Service which in turn sends it back to the  Evolutionary Operators Service for applying the genetic operators, namely selection, crossover and mutation.

{\textbf{DERs Service}}
The DERs Service provides other services dynamic and static data about the DER components. Examples of dynamic data are the actual state of batteries, forecasting data for the generation of RERs, consumption and market prices which are continuously changed according to different factors such as the weather. Static data encompasses the number and type of DER components,  technical constraints for the conventional energy resources e.g. minimum and maximum capacity, ramping limits and minimum up and down times, to name a few. Both types of data are stored in a database where each DER can insert and update its related data automatically, if it has an Energy Management System Interface (EMS-IF).  Otherwise, a manual insertion and update is required.  The Evolutionary Operators Service and the Interpretation Service instances need such data for the generation of the initial population and for the chromosome interpretation process as described later.  

{\textbf{Interpretation Service}}
As its name implies, it is responsible for interpreting the chromosomes in the context of  the optimization problem solved. For controlling DERs, the Evolutionary Operators Service generates scheduling operations represented by genes with relative values (e.g. in percent of the maximum providable power within a given time interval) representing the requested power share from each  DER at specific time interval. These values must be interpreted by converting them  to absolute values for evaluation (simulation) purposes. For example, for RERs, the relative generation values are multiplied by the corresponding forecasting data of the RERs to obtain the absolute values of a certain schedule. Since the interpretation process can require much computing  time according to the size of chromosomes, the framework  can  deploy as many Interpretation Service instances as required allowing a parallel interpretation  for  scalability. 

{\textbf{Calculation Service}}
The Calculation Service (or simulator) performs the calculations required to evaluate the individuals of the distributed population. It is called by the Interpretation Service for evaluating the offspring with respect to the given evaluation criteria. It takes a list of unevaluated individuals as the input and outputs the related evaluation results for each individual.

{\textbf{Container Management Service}}
The Container Management Service creates as many Interpretation Service  and Calculation Service instances as needed allowing runtime scalability. After creating and initializing the required instances successfully, the Container Management Service publishes a ready signal in order to start the processing of the optimization job.

{\textbf{Supporting and Learning Service}}
Typically, EAs start to generate the initial population randomly which ensures the necessary diversity of the start population and allows for an initial breadth search. On the other hand, using a given solution of a similar task can speed up the search at the risk of pushing the search into the direction of these solutions. Thus, only a few prior solutions should be taken as a part of the initial population. This can significantly accelerate an EA \cite{jakob2008fast}.  This service supports the Evolutionary Operators Service by generating the initial population and can use some already-found solutions (i.e. scheduling plans in case of DERs scheduling) for this based on predefined selection criteria.

{\textbf{Local Search Service}}
The Local Search Service is an extension of a deployed EA to support Memetic Algorithms (MAs). %This service provides the ability for using appropriate local search methods or heuristics to accelerate the evolutionary search of an EA by local improvement of the offspring. % While other solutions exist, for complex and multidimensional optimization problems MAs can be especially useful for finding an optimal solution quickly \cite{jakob2010MA}.

\begin{figure}[!h]
	\centering
	\includegraphics[width=0.845\textwidth]{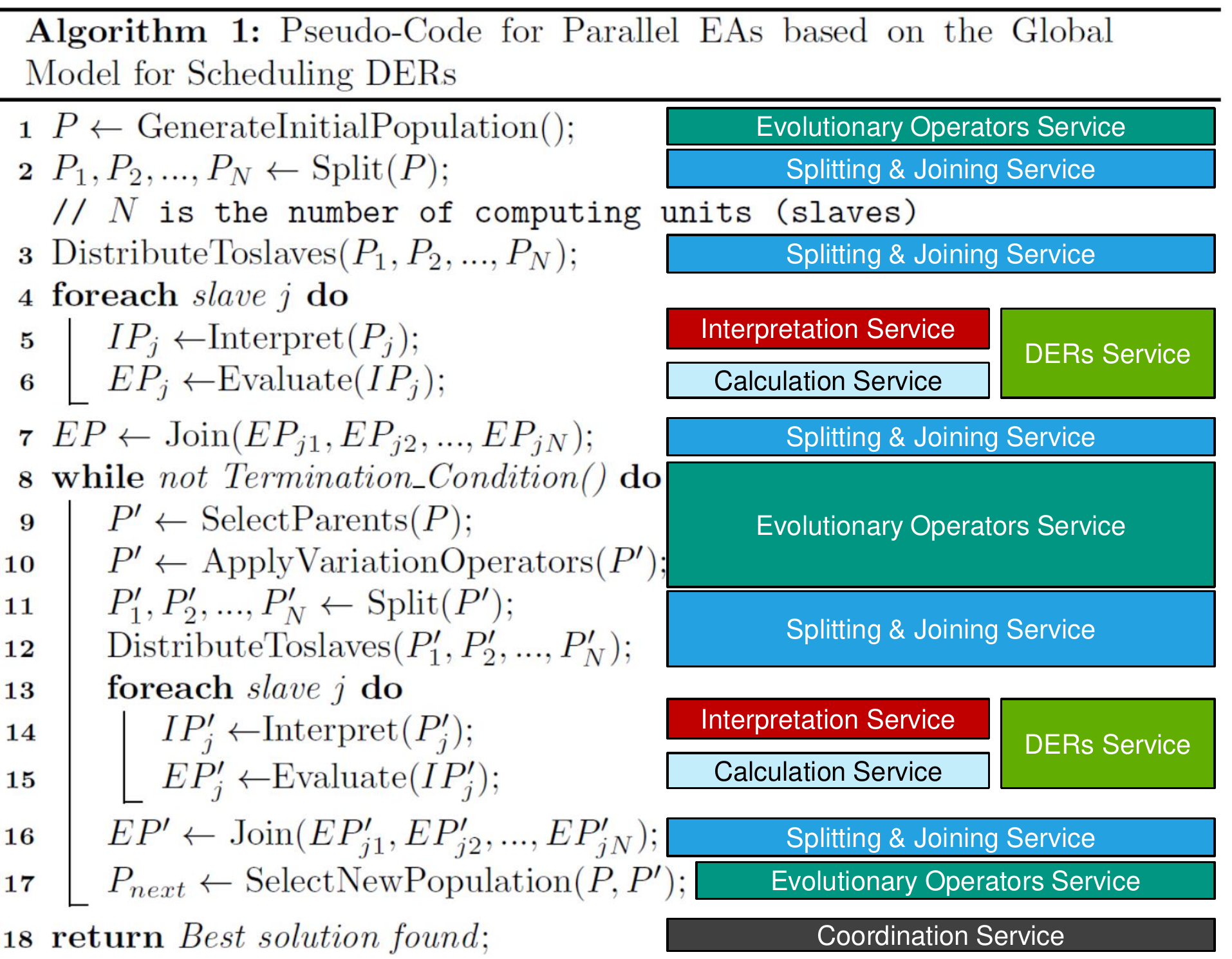}
	\caption{Mapping the related microservices to the pseudo-code of the parallel EAs based on the Global Model for scheduling DERs}
	\label{fig:pseudocode_mapping}
\end{figure}
The publish/subscribe pattern is used to realize the communications between the scalable microservices i.e. the Interpretation Service and the Calculation Service as well as between them and the other (non-scalable) microservices. The use of the publish/subscribe messaging paradigm ensures a seamless deployment, full decoupling among the services and an efficient and reliable data exchange among the services(cf. \cite{ref8PEA}). However, RESTful service APIs are useful for enabling the communication among the services specifically the microservices that are non-scalable in runtime, namely the Coordination Service, the Evolutionary Operators Service, the Optimization Task Coordination Service,  the Splitting \& Joining Service, the Container Management Service and the DERs Service. In Figure \ref{fig:pseudocode_mapping}, the pseudo-code of the parallel EAs --based on the Global Model-- for scheduling DERs is mapped to the related microservices.
\section{EA GLEAM for Scheduling DERs}
\label{sec:gleam}
The process of scheduling DERs consists of a set of scheduling operations that determine which DERs are involved in the power generation process and to what extent, in order to supply the required energy per time interval. The concrete EA GLEAM  \cite{ref1GLEAM} is integrated into the Evolutionary Operators Service for scheduling DERs, as it has proven its suitability for general scheduling problems in several different applications e.g. \cite{jakob2008fast}. GLEAM is acting as a master of the Global Model and generates the initial population, it applies the genetic operators and calculates the fitness value for each chromosome. 

The main feature that distinguishes  GLEAM from other EAs is its flexible coding used to optimize not only time-dependent processes but also any other optimization problems such as scheduling and design optimization. The coding in GLEAM is based on a set of  genes that are linked together forming a linear chain which represents a chromosome. The length of the chromosomes can either be fixed or altered dynamically by evolution.  In the following section, the GLEAM based solution for chromosome representation and interpretation for scheduling DERs is described.

\subsection{Solution Representation and Interpretation}

\begin{wrapfigure}{r}{0.52\textwidth}
	\includegraphics[width=0.52\textwidth, height= 2cm]{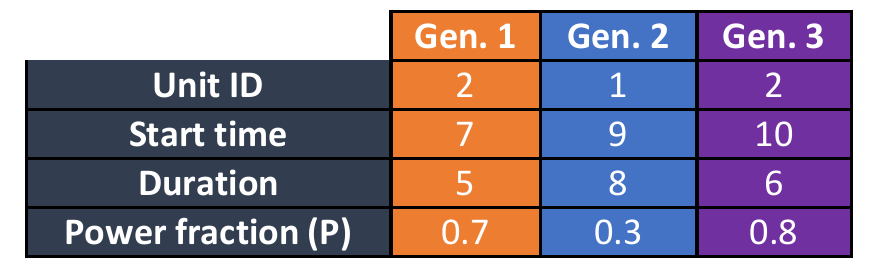}
	\caption{A chromosome with three genes encoding a possible solution to schedule two generation units}
	\label{fig:codingchromosome}
\end{wrapfigure}

Typically, a scheduling problem is broken down into several scheduling operations (e.g. one or more for each DER) which are represented by genes. In GLEAM, the structure of a gene  is flexible and the number and types of its decision parameters are defined related to the nature of the optimization problem. %In a simple case, the structure of each gene can code only one parameter. However, in more complicated cases, it can be expanded to code more complex problem with several integer and real parameters.
The genes are moved as a whole by the respective genetic operators, which corresponds directly to the change in the sequence of the planning operations. Each scheduling operation is coded by one gene that consists of a fixed gene ID, which corresponds to the unit ID of the related DER, and the following decision variables: start time, duration and the power fraction as shown in Figure \ref{fig:codingchromosome}. 
While the start time  is used to determine the start time of taking energy from this DER and the duration refers to the number of time intervals to which this setting applies, the power fraction variable determines the amount of energy in relation to the forecasted  maximum that can be obtained from a DER.
Since the number of required scheduling operations is not known a priori, the length of each chromosome is changed  dynamically by the evolution. Mutation operators such as the duplication, deletion or insertion of individual genes or gene segments are used to alter the length of chromosomes (cf. \cite{ref1GLEAM,ref4GLEAM} for a detailed discussion).

\subsubsection{Chromosome Interpretation}

For the construction of an allocation matrix, the genes of a chromosome are successively treated so that a later gene overwrites matrix entries of the previous ones with the same Unit ID. This is considered as the first step of chromosome interpretation by the Interpretation Service. For each chromosome list, the first task is generating an allocation matrix where the number of rows $m$ is equal  to the number of resources, i.e. DERs in this chromosome list, and the number of columns $n$ represents the time intervals. When the building of the allocation matrix is finished, the Interpretation Service starts the second step of interpretation, namely converting the relative values of power fraction to absolute values by multiplying each value in the allocation matrix by the corresponding values of the actual maximum power generation supplied by the DERs Service for the corresponding time interval. As a result, a new matrix with absolute values is produced and prepared  for evaluation (simulation) by the Calculation Service.
%\begin{wrapfigure}{r}{0.7\textwidth}
%	\includegraphics[width=0.7\textwidth]{fig/5.pdf}
%	\caption{Interpretation of one chromosome with three genes for scheduling 2 DERs and 24 one-hour intervals}
%	\label{fig:allocationmatrix}
%\end{wrapfigure}  
%Figure \ref{fig:allocationmatrix} shows an example of the creation of an  allocation matrix for a chromosome with three genes where the first and the third genes have the same Unit ID as shown in Figure \ref{fig:codingchromosome}. The third gene overwrites the first one at the time intervals from 10 to 15 as shown in Figure \ref{fig:allocationmatrix}. 
%\begin{figure}[!h]
%	\centering
%	\includegraphics[width=0.5\textwidth]{fig/5.pdf}
%	\caption{Interpretation of one chromosome with three genes for scheduling 2 DERs and 24 one-hour intervals}
%	\label{fig:allocationmatrix}
%\end{figure}

\section{Evaluation}
\label{sec:eva}
In this section, the performance of the proposed distributed solution with respect to scalability is discussed.
First, three use case scenarios are introduced in section \ref{sec:usecase}. Afterwards, the mathematical optimization problem with objective functions and constraints is formulated. Thereafter, the GLEAM configuration and the deployment of  the experiment using services on a cluster are described. The interpretation of the results will then be discussed in section \ref{sec:results}.

\subsection{Use Case Scenarios}
\label{sec:usecase}
\begin{figure}[!h]
	\centering
	\includegraphics[width=0.85\textwidth]{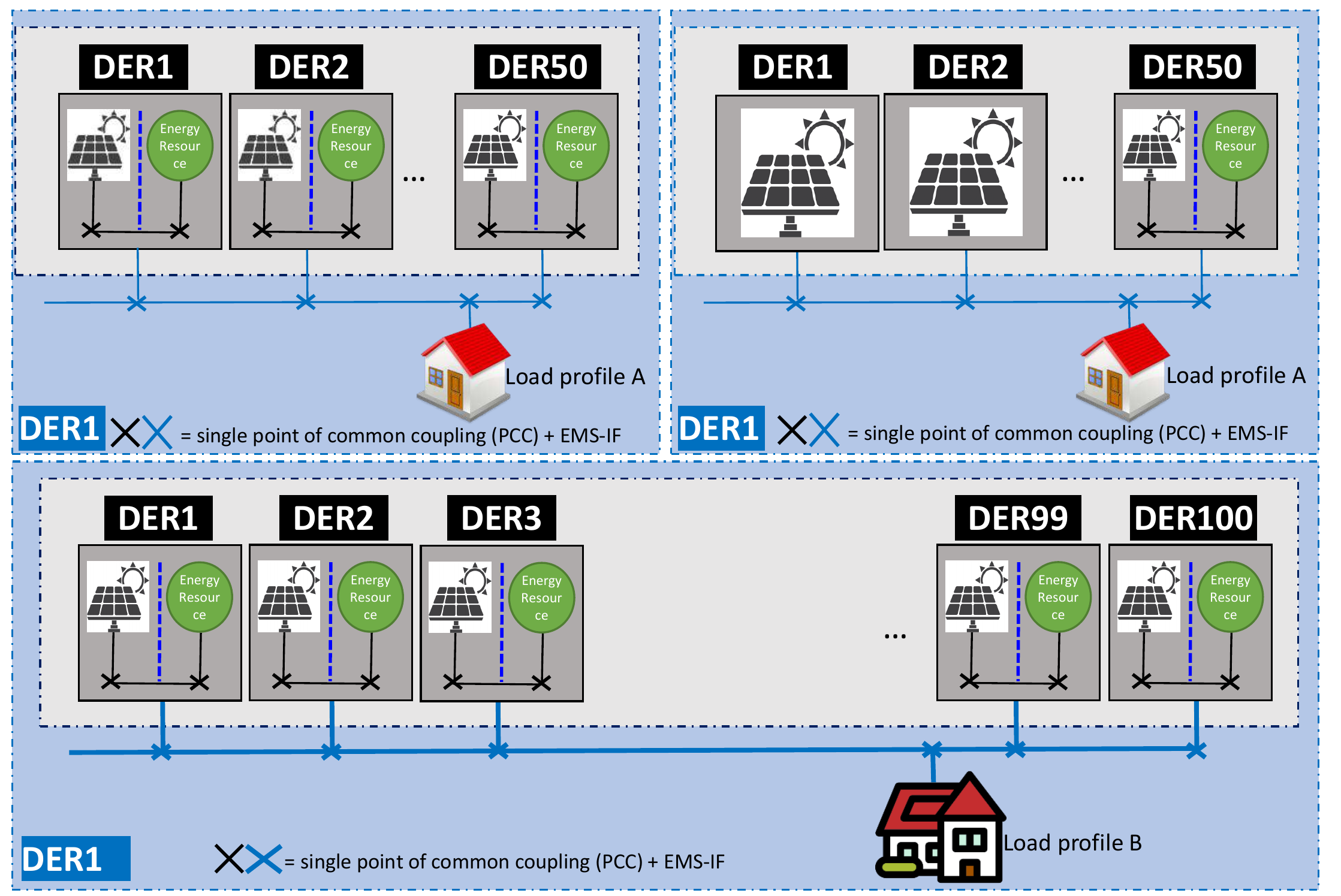}
	\caption{Use Case Scenarios used for evaluation}
	\label{fig:usecases}
\end{figure}

For evaluating the scalability and generality of the proposed approach, three DER scheduling scenarios instrumenting a different number of DERs and DER mixes (only PV, PV with other generation sources or storage) with predefined generation behaviour, and two different load profiles are defined, see Figure \ref{fig:usecases}. 
For defining renewable generation behaviour, the hourly real power generation data for 50 and 100 PVs provided by Ausgrid \cite{ref1dataset} is used.
Each DER has an EMS which manages and coordinates this DER. The EMS has a communication interface (EMS-IF) which provides flexibility of the DER in terms of the amount of energy that can be sold at a specific time interval with a specific price to consumers, e.g. the aggregated more or less controllable load (house symbols) as shown in Figure \ref{fig:usecases}. 

\begin{figure}[!h]
		\centering
	\includegraphics[height= 5cm]{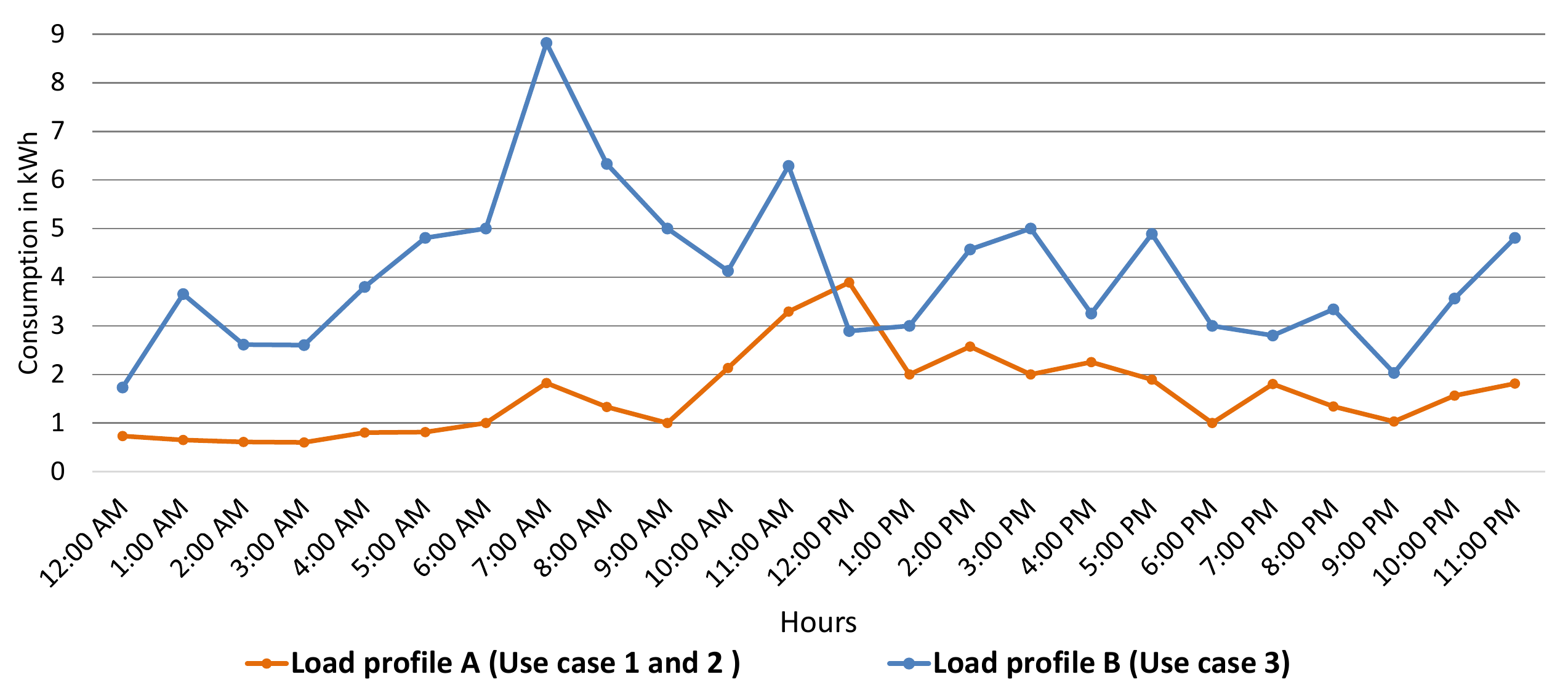}
	\caption{The two load profiles used for evaluation}
	\label{fig:consumption}
\end{figure}

In the first scenario depicted in the upper left part of Figure \ref{fig:usecases},  50 DERs can offer  power for 24 hours to cover a simple load profile (load profile A) as shown in Figure \ref{fig:consumption}. For the period between 7 and 17 o'clock, the EMSs offer the power to be sold from PVs and outside this period from other resources such as batteries or wind turbines. In the second scenario depicted in the upper right part of Figure \ref{fig:usecases}, the same load profile (load profile A) as in the first scenario is used. However, only a part of DERs, namely 25 DERs can offer power for the consumer 24 hours from the PVs combined with other resources. The other 25 DERs have only PVs and therefore can offer power only for 10 hours between 7 and 17 o'clock.  In the third  use case scenario depicted in the lower part of Figure \ref{fig:usecases}, 100 DERs provide the requested power for 24 hours to cover a more complex load profile (load profile B) as shown by the blue line in Figure \ref{fig:consumption}.
The main task of the distributed GLEAM is to minimize the daily bill costs of the customer by generating the optimal hourly scheduling plan for one day ahead. Additionally, there are some constraints which have to be fulfilled.
\vspace{-5mm}
\subsection{Objective Functions and Constraints}
%\begin{figure}[!h]
%	\subfloat[\tiny{Cost criterion}\label{fig:cost}]{\includegraphics[width=0.34\textwidth,height= 4cm]{fig/8_1.pdf}}
%	\subfloat[\tiny{Daily Total Deviation (DTD) criterion}\label{fig:daily}]{\includegraphics[width=0.34\textwidth,height= 4cm]{fig/8_2.pdf}}
%	\subfloat[\tiny{Penalty function for the Hours of Undersupply (HU)  criterion}\label{fig:penalty}]{\includegraphics[width=0.34\textwidth,height= 4cm]{fig/8_3.pdf}}
%	\caption{Mapping objective functions to the fitness function}
%	\label{fig:overhead}
%\end{figure} 
For the present evaluation, the Cost-Effective Operation Mode \cite{ref1EVA} is considered. 
Equation (\ref{eq:cost}) defines the cost function as a nonlinear (e.g. quadratic) function to be minimized for the above three use cases. 
\begin{equation}
\label{eq:cost}
Cost =  {\sum_{i=1}^{N}\sum_{t=1}^{T} C_{i,t}*(P_{i,t})} =  {\sum_{i=1}^{N}\sum_{t=1}^{T} [\alpha P^2_{i,t}+\beta P_{i,t}+ \gamma ]}
\end{equation}
where  $N$ is the number of DERs, $T$ is the number of the time intervals, $C_{i,t}$ is the price in (EUR) for each kWh taken from resource $i$ in time interval $t$, $P_{i,t}$ is the scheduled power in kWh taken from resource $i$ in time interval $t$  and $\alpha$, $\beta$ and $\gamma$  are the cost function coefficients defined for each DER at every time interval $t$.
Since DERs should work as much as possible by only using locally supplied power, the power balance within each DER is considered an important optimization objective. For achieving such balance, an additional objective function, namely the Daily Total Deviation (DTD) function shown in Equation (\ref{eq:balnace1}) is defined. It is the sum of absolute differences between the required power and the scheduled one at every time interval $t$. For arriving at a local balance DTD should be as low as possible.
\begin{equation}
\label{eq:balnace1}
DTD =  {\sum_{t=1}^{T}\left |\sum_{i=1}^{N}P_{i,t} - D_{t}\right |}
\end{equation}
where $D_{t}$ is the requested power by the load in time interval $t$ in kWh.
To guarantee that the evolutionary search process preferably finds solutions without undersupply at each hour, the Hours of Undersupply (HU) function shown in Equation (\ref{eq:balnace2}) is defined. It represents the number of hours of undersupply and takes an integer value between zero (the optimal case no undersupply) and $T$ (the worst case there are undersupply in all hours). The initial value of HU is zero.
\begin{equation}
\label{eq:HU}
HU = \left\{
\begin{matrix}
HU{++},& if\: D_{t} > \sum_{i=1}^{N}P_{i,t} : t \in (1,..,T) \\ 
HU & otherwise
\end{matrix}
\right.
\end{equation}
Due to the nonlinearity of the cost and DTD functions, the optimization problem is formulated as a nonconvex mixed-integer nonlinear optimization problem. Moreover, it is a multi-objective problem: % As shown in Equations (\ref{eq:optimProb}) and (\ref {eq:balnace2}), the cost and DTD functions are considered as objective functions and HU as an equality constraint which should be zero.
\begin{equation}
\label{eq:optimProb}
Minimize[Cost , DTD]
\end{equation}
subject to
\begin{equation}
\label{eq:balnace2}
HU =  0
\end{equation}
The optimization problem defined above in Equations (\ref{eq:optimProb}) and (\ref {eq:balnace2}) is an adequate problem for our evaluation, since the scheduling of DERs is NP-hard optimization problem \cite{NpHardDER} and formulated as nonconvex mixed-integer nonlinear optimization problem which need lots of computational power. Moreover, the numerical solution for such optimizarion problem is typically too complex for exact optimization methods \cite{NpHard}. Hence, EAs represent a robust and powerful alternatives \cite{NpHardEA}. The EA GLEAM should minimize the cost and DTD objective functions as far as possible while holding the constraint HU. 

The Calculation Service is responsible for computing the values of the above objective functions (criteria) and constraint for each individual i.e. chromosome. %For giving GLEAM a hint into what direction the populations should evolve, GLEAM uses (as other multiobjective solutions also) a weighting function called fitness curve which needs to be defined in the GLEAM configuration. 
The weighted sum defined in Equation (\ref{eq:weightedsum}) is used to combine the results of the  criteria into a fitness value. %For this purpose, the numerical value provided by the Calculation Services for each criterion must be mapped to a uniform fitness scale using one of standard normalization functions of GLEAM, namely linear, exponential and mixed linear-exponential. 
The fitness scale is set to a range between ``0'' and ``100.000''. The fitness value determines in GLEAM the likelihood of an individual reproducing and passing on its genetic information. This happens especially when choosing a partner and deciding whether to accept or reject the offspring when forming the next generation. 
%For the cost and DTD criteria, the inversely proportional exponential function is used as shown in Figures \ref{fig:cost} and \ref{fig:daily}, respectively.
In order to handle the equality constraint HU, a Penalty Function $PF$ shown in Equation (\ref{eq:PF}), which yields a value between zero and one (no undersupply) is defined. The fitness determined from the other two criteria is multiplied by this, so that an undersupply of 5 hours already reduces the fitness value to a third.
\begin{equation}
\label{eq:PF}
PF(HU) = (1- \frac{1}{T}HU) 
\end{equation}
\begin{equation}
\label{eq:weightedsum}
Fitness = (0.4*Cost + 0.6*DTD)*PF(HU) 
\end{equation}

\subsection{Deployment on a Cluster}
For instrumenting the solution on a  computer cluster, it is deployed on a cluster with four computing nodes where each node has 32 Intel cores (2,4 GHz) resulting in 128 independent computing units, 128 GB RAM and an SSD disk. The nodes are connected to each other by a LAN with 10 GBit/s bandwidth. A modern software environment based on container automation technology guarantees a seamless deployment of the microservices on the cluster. For enabling containerization, the most popular open source software, namely Docker\footnote{www.docker.com} is used. Docker performs operating-system-level virtualization to isolate the applications. This is achieved by running containers on the Docker engine that separates the applications from the underlying host operating system. For container orchestration, Kubernetes\footnote{www.kubernetes.io} is chosen as  container orchestration system. 
\begin{figure*}[!h]
	\centering
	\includegraphics[width=0.85\textwidth, height= 6cm]{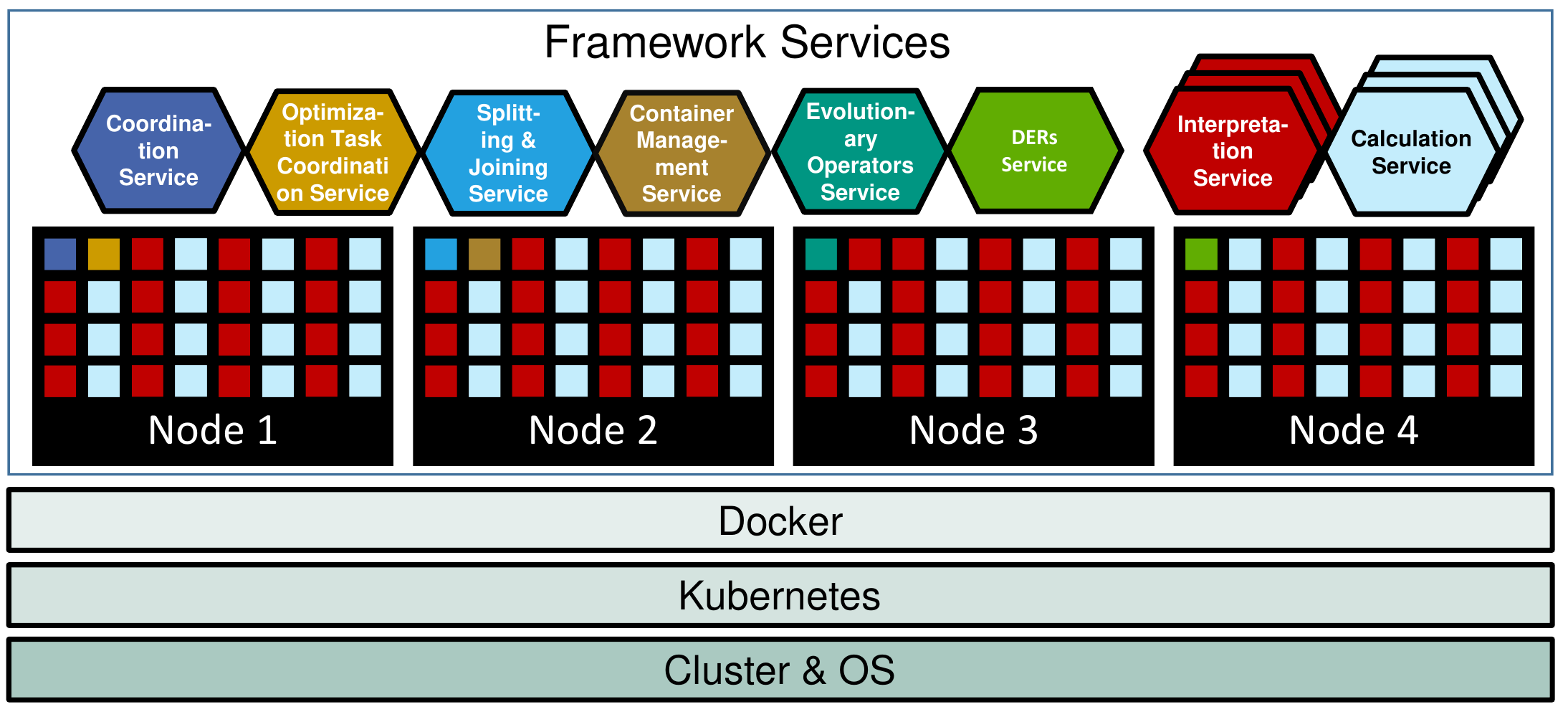}
	\caption{Mapping the proposed microservice architecture to the cluster with four nodes}
	\label{fig:clusternode}
\end{figure*}
It is used in many production environments due to its flexibility and reliability. Kubernetes defines several building blocks which are called Pods to separate ``computing loads" from each other and provide mechanisms to deploy, maintain and scale applications. A Pod is the smallest building block in the Kubernetes object model and represents one or more  running processes. The highly distributable Redis\footnote{www.redis.io} is deployed  as an in-memory database serving as  a temporary storage for intermediate results and their exchange. Redis provides the publish/subscribe messaging paradigm. The persistent database storing DER forecasting data for power generation, power consumption and market prices is implemented by using the InfluxDB\footnote{www.influxdata.com} time series database.
Figure \ref{fig:clusternode} shows the technological layers and an example of how the services can be mapped to the CPUs on the four nodes. %While the Coordination Service and the Optimization Task Coordination Service are deployed on the first node, the Splitting \& Joining Service and Container Management Service are deployed on the second node. The Evolutionary Operators Service is running on the third node and the DERs Service is running on the last node. 
It is important to notice that the required instances from the Interpretation Service and the Calculation Service are distributed over all nodes dynamically.

\subsection{Results and Discussion}
\label{sec:results}
In the following, the efficiency of the parallel method for scheduling DERs developed based on modern technologies, namely microservice and container virtualization is introduced. The achieved quality of the schedules using distributed EAs and the scalability in cluster environments are particularly discussed. 

\subsubsection{Resulting Schedules}
\begin{figure}
	\centering
	\subfloat[The optimal scheduling plan for use case 1\label{fig:planus1}]{\includegraphics[width=.99\textwidth, height=2.2cm]{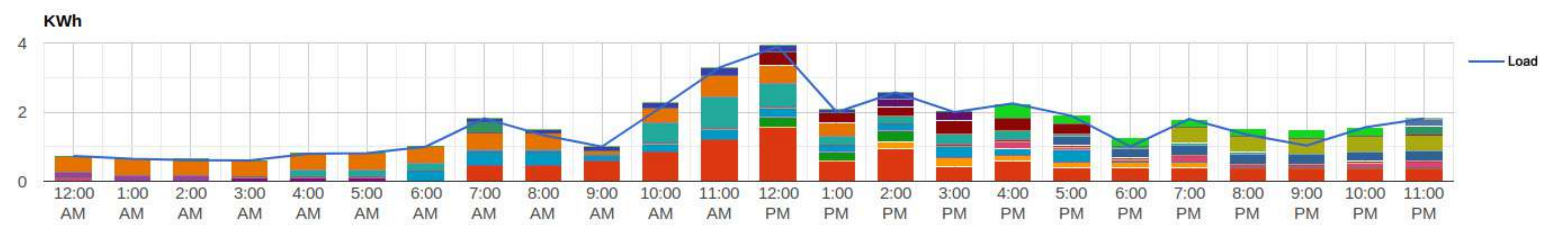}}\\
	\subfloat[The optimal scheduling plan for use case 2\label{fig:planus2}]{\includegraphics[width=.99\textwidth, height=2.2cm]{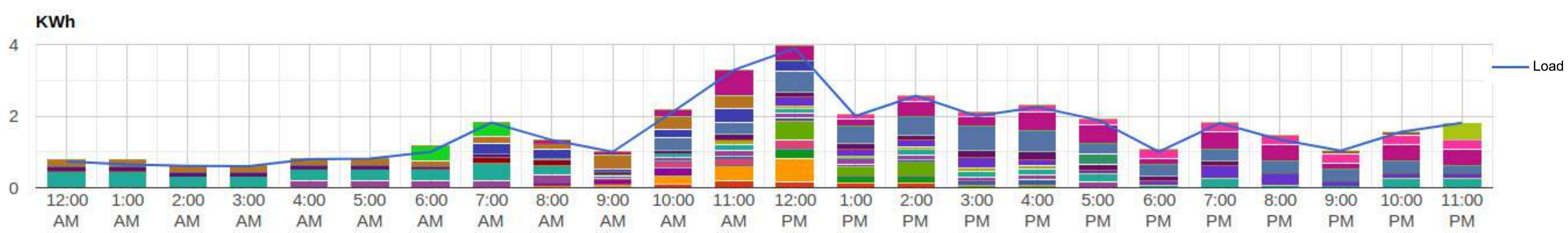}}\\
	\subfloat[Proportion of energy generated by PV in use case 2\label{fig:Usecase2_25PVs}]{\includegraphics[width=.99\textwidth, height=2.6cm]{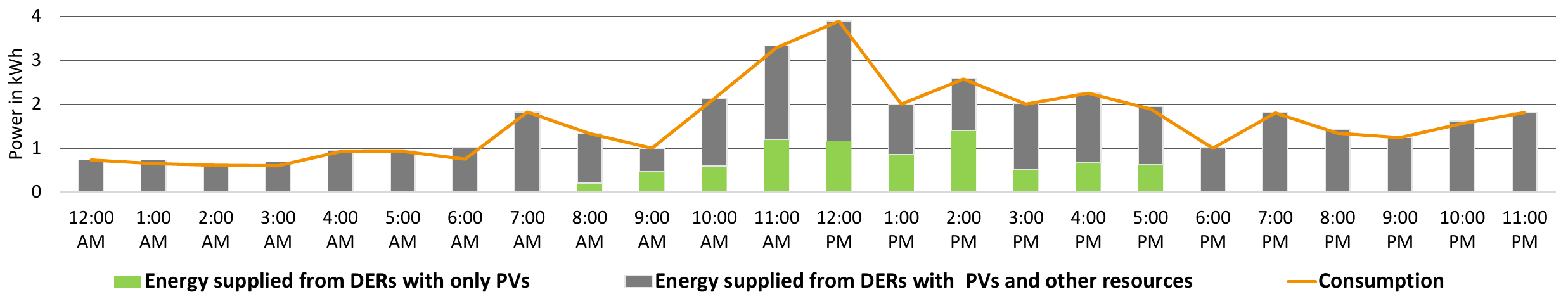}}	\\
	\subfloat[The optimal scheduling plan for use case 3\label{fig:planus3}]{\includegraphics[width=.99\textwidth, height=2.2cm]{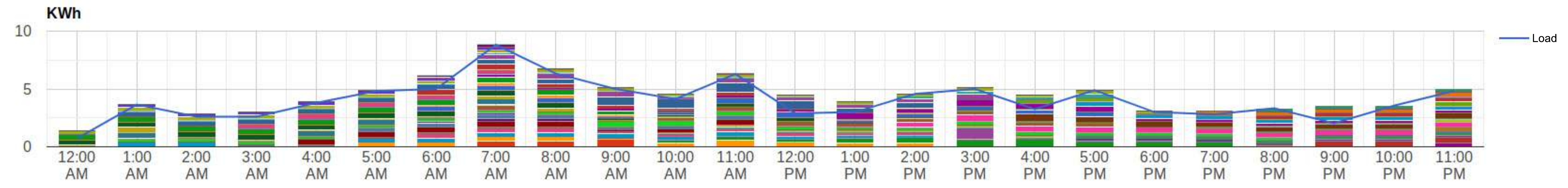}}
	\caption{The optimal scheduling plans obtained for the defined use cases, coloured rectangles represent the amount of scheduled power taken from the DERs contributed in the schedule, blue line is the consumption profile}
	\label{fig:planusecases}
\end{figure}
For achieving a good trade-off between exploration and exploitation, appropriate strategy parameters of the EA, namely the size of the population and the number of generations must be determined. For this, we perform several tests with 120 slaves and varying the population size as follows: 120, 180, 240, 300 and 420 individuals,  so that each slave at minimum can process one individual. The number of offsprings per pairing is set to eight.
To limit the effort, the number of generations is set to 420. For the first use case an optimal schedule with 21 DERs from the available 50  DERs--depicted in Figure \ref{fig:planus1}--, can be obtained with a population size of 180 individuals. For the second use case, GLEAM needs more individuals, namely 300 to explore the search space sufficiently and to find an optimal scheduling plan --shown in Figure \ref{fig:planus2}--  using more DERs, namely 31 from the available 50  DERs. In comparison with the first use case, the number of scheduling operations (genes) and the corresponding number of evaluations are increased significantly. This is due to the fact that one half of the used 50 DERs are restricted to supply power only for 10 hours per day resulting in a more heterogeneous search space and a further constraint to GLEAM.
Figure \ref{fig:Usecase2_25PVs} shows how the required energy is covered by the two types of DERs considered in the second use case. As shown, the pure DERs with only PVs contribute with a generation portion between 13\% (at 8 o'clock) and 54\% (at 14 o'clock).
For the third use case with 100 DERs and a more complex load profile, a scheduling plan --shown in Figure \ref{fig:planus3}-- is found with a population of size 240 individuals. %It is noteworthy that in the use cases 1 and 3, the number of evaluations increases by only 69\% if the number of considered DERs are doubled. 

\subsubsection{Framework Scalability}

In order to assess the performance of the proposed software architecture, we analyze the scalability of the framework for the above three use case scenarios introduced in section \ref{sec:usecase}. The number of computing units (slaves), namely the Interpretation Service instances as well as the number of the corresponding Calculation Service instances is varied between 1 and 120 so that the minimum of two cores is left on each node for the OS.
 \begin{table}[!h]
	\centering
	\begin{tabular}{@{}cccc@{}}
		\toprule
		\multicolumn{1}{c}{}           & \multicolumn{3}{c}{\textbf{Computational Time in Minutes}}        \\ \cline{2-4}
		\textbf{\#of computing  units} & \textbf{Use case 1} & \textbf{Use case 2} & \textbf{Use case 3} \\ \midrule
		\textbf{1}                     & 780                 & 1290                 & 4175                \\
		\textbf{8}                     & 133                 & 237                 & 611                 \\
		\textbf{24}                    & 67                  & 123                 & 342                 \\
		\textbf{40}                    & 55                  & 99                  & 275                 \\
		\textbf{56}                    & 56                  & 97                  & 270                 \\
		\textbf{72}                    & 54                  & 90                  & 263                 \\
		\textbf{88}                    & 55                  & 85                  & 255                 \\
		\textbf{104}                   & 44                  & 80                  & 250                 \\
		\textbf{112}                   & 43                  & 72                  & 246                 \\
		\textbf{120}                   & 38                  & 66                  & 200                 \\ \bottomrule
	\end{tabular}
	\caption{The computational time of the three use cases when increasing the number of computing  units (slaves)}
	\label{tab:comptime}
\end{table}
\vspace{-10mm}

Table \ref{tab:comptime} shows the scalability results of the three use cases where the total time for each optimization job is measured. It can be concluded that by increasing the difficulty of the optimization problem, the total time needed to find an optimal solution is increased. 
Therefore, the scheduling process for the second  and third use cases takes more time as the first one, since GLEAM performs more evaluations. Within 420 generations, GLEAM achieves 548566 evaluations for the first use cases, 919908 for the second one and 799462 for the third use case. By using more computing units,  the framework is able to reduce the total time from 780 to 38 minutes in the first use case, from  1290 to 66 minutes in the second use case and from 4175 to 200 minutes in the last use case. For each use case, the computation time of the parallel implementation decreases more slowly at a certain point, since the communication overhead of the increased number of computing units (slaves) exceeds the increased performance of the parallelization. 

%\begin{figure*}[!h]
%	\subfloat[\tiny{Computation time needed in use case 1 vs. use case 2}\label{fig:speedup1}]{\includegraphics[width=0.5\textwidth]{fig/SpeedUp_1and2.pdf}}
%	\subfloat[\tiny{Computation time needed in use case 1 vs. use case 3}\label{fig:speedup2}]{\includegraphics[width=0.5\textwidth]{fig/SpeedUp_1and3.pdf}}
%	\caption{The optimal scheduling plans obtained for the defined use cases}
%	\label{fig:comptime}
%\end{figure*}   
%\begin{table}[!ht]

%\begin{tabular}{@{}cccc@{}}
%	\toprule
%	\multicolumn{1}{l}{\textbf{Use case}} & \textbf{\#of individuals} & \textbf{\#of generations} & \textbf{Fitness value} \\ \midrule
%	1                                     & 120                       & 300                       & 94074                  \\
%	2                                     & 120    & 300      & 94587   \\
%	3                                     & 240                       & 420                       & 59154                  \\ \bottomrule
%\end{tabular}
%	\label{tab:exp}
%	\caption{EA tests to trade off exploration and exploitation}
%\end{table}

\section{Conclusion and Future Work}
\label{sec:cons}
In this paper, a new parallel, highly modular, flexible and scalable method for scheduling Distributed Energy Resources (DERs) based on Evolutionary Algorithms (EAs) is presented. In contrast to other optimization methods, the new proposed solution enables an efficient parallelization of EAs, full runtime automation and an easy deployment on high performance computing environments such as clusters or cloud environments. Furthermore, it provides the ability to deal with a comparably large number of DERs.  Modern software technologies, namely microservices, container virtualization and the publish/subscribe messaging paradigm are exploited to develop the desired method. The architecture clearly separates functionalities related to EAs and the ones related to scheduling DERs. For each functionality, a microservice is designed and implemented. Furthermore,  container virtualization is utilized to automatically deploy the microservices on nodes of an underlying cluster to perform their tasks.  The combination of microservices and container virtualization enables an easy integration of an existing EA into the framework and facilitates the communication with other required services like simulators and forecasting tools for power generation and consumption, market price and weather. Furthermore, using  the publish/subscribe messaging paradigm guarantees a seamless data exchange between the scalable services which are deployed on-demand. 

In order to evaluate the functionalities of the proposed solution, three use case scenarios with different types and numbers of DERs are defined and studied. The scalability of the framework is demonstrated by varying the number of computing units between 1 and 120. The results show that the new distributed solution is an applicable approach for scheduling a scalable number of DERs using EAs based on the mentioned three lightweight technologies in a scalable runtime environment.

As part of future work, more detailed evaluations related to the communication overhead of the solution will be undertaken. Other parallelization models for EA such as Coarse-Grained Model can also be applied and compared with the current presented approach.  Furthermore, a comparison with a central solver like simplex or other distributed population-based metaheuristics will be considered. %The Supporting and Learning Service mentioned could be implemented to accelerate the EA based on learned knowledge about prior solutions. 


\bibliographystyle{bibtex/splncs03}
\bibliography{FTC2020}

\end{document}